 \documentclass[preprint2]{aastex}

\slugcomment{Astrophysical Journal}

\shorttitle{Similarity Properties and  Scaling Laws of Radiation Hydrodynamic Flows}
\shortauthors{Falize \'E. et al.}

\usepackage{color}
\begin{document}


\title{Similarity Properties and  Scaling Laws of Radiation Hydrodynamic Flows in Laboratory Astrophysics}

\author{\'E. Falize\altaffilmark{1,2}, C. Michaut\altaffilmark{2}  \& S. Bouquet\altaffilmark{1,2} }

\email{emeric.falize@cea.fr}

\altaffiltext{1}{CEA-DAM-DIF, F-91297 Arpajon, France}
\altaffiltext{2}{Laboratoire Univers et Th\'eories (LUTH), Observatoire de Paris, CNRS, Universit\'e Paris-Diderot, 92190 Meudon, France}

\begin{abstract}
The spectacular recent development of modern high-energy density laboratory facilities which concentrate more and more energy in millimetric volumes allows the astrophysical community to reproduce and to explore, in millimeter-scale targets and during very short times, astrophysical phenomena where radiation and matter are strongly coupled. The astrophysical relevance of these experiments can be checked from the similarity properties and especially scaling laws establishment, which constitutes the keystone of laboratory astrophysics. From the radiating optically thin regime to the so-called optically thick radiative pressure regime, we present in this paper, for the first time, a complete analysis of the main radiating regimes that we encountered in laboratory astrophysics with the same formalism based on the Lie-group theory. The use of the Lie group method appears as systematic which allows to construct easily and orderly the scaling laws of a given problem. This powerful tool permits to unify the recent major advances on scaling laws and to identify new similarity concepts that we discuss in this paper and which opens important applications for the present and the future laboratory astrophysics experiments. All these results enable to demonstrate theoretically that astrophysical phenomena in such radiating regimes can be explored experimentally thanks to powerful facilities. Consequently the results presented here are a fundamental tool for the high-energy density laboratory astrophysics community in order to quantify the astrophysics relevance and justify laser experiments. Moreover, relying on the Lie-group theory, this paper constitutes the starting point of any analysis of the self-similar dynamics of radiating fluids. 
\end{abstract}

\newpage
\section{Introduction}
Modern high-energy density facilities (including powerful lasers and Z-pinch machines), which concentrate more and more energy in millimetric volumes, allow to bring up the matter, reproducibly, to new extreme states of density, temperature and velocity in laboratory \citep{Drake06, Remington06, Moses09}. These new experiment classes allow to characterize and measure the fundamental properties of matter in new physical regimes. Thanks to this new experimental capability, various hydrodynamical flows with an astrophysical interest, such as high Mach number flows \citep{Loupias07, Hartigan09, Gregory10a} or hydrodynamical instabilities \citep{Drake05} such as Rayleigh-Taylor \citep{Kuranz09} or recently Kelvin-Helmholtz instabilities \citep{Hurricane09, Harding09}, have been studied. With the flexibility of these experiments we can examine and diagnose the complex static or dynamic interaction of matter with an external magnetic field or/and radiation. Using adapted target design, it is possible to create intense radiation which drives the flows such as X-ray thermal waves \citep{Back00a, Back00b} as well as intense hydrodynamics flows, which leads to the radiation of plasmas as radiative shock waves \citep{Bouquet04, Koenig06, Michaut07, Michaut09}. Thus, the powerful facilities provide a unique opportunity to make progress in the understanding of these extreme phenomena which had never been created before at laboratory scales but they are very common in high-energy astrophysics environments.  
The experimental challenge is to create and to maintain a laboratory system which is similar to its astrophysical counterpart. Thus, the fundamental problem of laboratory astrophysics is to determine the astrophysical relevance of these experiments and to reconcile the spatial and temporal scales which are so different as well as the thermodynamical regimes. It is only by a rigorous and detailed study of the scalability properties of such flows and the scaling laws establishment that we can determine the possibility of reproducing the astrophysical phenomena. The laboratory experiments provide key insights into our understanding of these phenomena at astrophysical scales which is generally partial because of the difficulty of observing them. Thus, the possibility to use an adapted scaling law in order to reproduce, at diagnosis scales, high-energy astrophysical phenomena appears as an essential complement in order to test the astrophysical models and simulations. Beyond their interest in laboratory astrophysics, the scaling laws play a crucial role in all high-energy density physics since they can be used in order to adapt a target design from a powerful facilities to another. Moreover they can consist in a powerful tool for numerical simulations.\\
Several theoretical studies of similarity properties and scaling laws have been published in purely hydrodynamic regimes \citep{Basko98, Ryutov99, Ryutov02, Ryutov03} and in ideal MHD \citep{Ryutov00, Ryutov01}. Concerning radiation hydrodynamic regimes, few studies have been published. In optically thin radiating plasma regimes, only the similarity properties have been considered \citep{Ryutov01,Castor07}. For the optically thick regime, \citet{Murakami02} have studied scaling laws in inertial fusion context for specific flow classes. Recently the scalability of two-temperature regime (electron and ion temperatures) has been considered \citep{Falize10}.\\		 
This paper consists in an exhaustive study of similarity properties and scaling laws of radiation hydrodynamic flows in different regimes which are or will be achieved in laboratory with current or future facilities. For each regime, connections to astrophysical objects and phenomena are discussed. We have based our analysis on an original approach with Lie group symmetries. This powerful formalism appears as a systematic method which provides easily and systematically the establishment of the scaling laws of a given problem. Although the scaling laws can be obtained by classical dimensional analysis formalism, it is only by the Lie group symmetry that the different invariance concepts can be introduced rigorously. A new similarity concept, the \emph{global invariance} \citep{Falize08, Falize09b}, which introduces important perspectives in laboratory astrophysics, is presented in this paper. The latter is organized as follows: firstly we present an extended classification of similarity concepts that we use in laboratory astrophysics; secondly we examine the scalability properties of optically thin radiation hydrodynamic flows. Finally, before concluding, the scaling laws and the similarity properties of optically thick radiating fluids are considered in two specific diffusion regimes including the regime where density energy and pressure of radiation are not negligible compared to the same matter quantities.

\section{Scaling invariance concepts and similarity experiments}
 Although a profound connexion exists between the scalability properties of flows and their self-similar behaviors, it is important to separate these two distinct concepts. Indeed it is crucial to bear in mind than two similar flows have not necessary a self-similar dynamics. In order to illustrate this point, let's consider the example of the important high-energy density phenomenon of X-ray radiative heating of opaque material. Its physical phenomenology is described and studied by \citet{Pakula85,Kaiser89}. During the first moments, a decelerate supersonic Marshak wave propagates into the opaque matter. When the radiative front becomes subsonic, a shock is formed in the head of the wave which leads to the classical structure of ablative wave. In this physical situation, two similar flows can be defined, since in specific cases, scaling laws can be established for this complex phenomenon, whereas its global dynamics is clearly non-self-similar. \newline Thus, generally,
two flows are called to be similar when there exists a transformation group which allows to pass continuously from the laboratory plasmas to astrophysical phenomena. A great variety of non trivial transformations can agree with this general definition. For instance in hydrodynamics systems, \citet{Drury00} proposed using the projective symmetry in order to reproduce a supernova explosion by an implosion of target.  This approach has been extended to optically thin radiating fluid dynamics by \citet{Falize08}.\\
In this paper the homothetic symmetry only is examined in detail. 
Although the dimensional analysis allows to obtain the similarity properties \citep{Sedov59} of the physical system and to establish the scaling laws, we favor the one-parameter homothetic Lie group \citep{Birkhoff50,Bluman74}. Thanks to this group, the connection between the astrophysical ($X_{i}$) and the laboratory ($\tilde{X}_{i}$) quantities are defined by the general transformation:
\begin{equation}
X_{i}=\lambda^{\delta_{i}}\tilde{X}_{i},
\end{equation}
where $\lambda$ is the group parameter and $\delta_{i}$'s are the homothetic exponents of the rescaled quantities.  
Although we do not discuss the problem of the rescaling of the initial and boundary conditions, it is trivial that the latter must be invariant in all the scale transformations that we will discuss and establish in this paper. This intuitive but constraining condition is discussed in detail in \citet{Ryutov99,Ryutov00}. \newline
The important work realized since a few decades on scalability of laboratory flows have permitted to introduce new similarity concepts in order to define laboratory experiments.  
The first important invariance concept is the \emph{perfect similarity} which has been introduced by \citet{Ryutov03}. The authors pointed out that in hydrodynamic and non dissipative MHD systems a simple transformation exists and consists in rescaling only the spatial and temporal coordinates ($r=A\times \tilde{r}$, $t=A\times \tilde{t}$ where A is a free parameter). In these physical regimes, this scaling law requires no approximation of equations of state (EOS), which makes a very attractive invariance notion especially when the knowledge of these informations are poor. Nevertheless it can not be generally used in radiation hydrodynamic systems and more general invariance concepts must be introduced where all physical quantities are rescaled.\\
Two distinct similarity concepts must be introduced \citep{Falize08, Falize09b}: the \textit{absolute similarity} which consists in the rescaling of all physical quantities and leaves invariant the equations and the \textit{global similarity} which is a more general framework and is justified by the Lie group theory. In the latter case, only the form of equations is invariant and  the different ionization rates or external physical fields such as magnetic fields \citep{Falize09} are absorbed in the scaling laws form. 
Let's note in general the laboratory plasmas are composed by species with more important atomic weight than in astrophysical situations due to some technological limitations. Thanks to the \emph{global similarity}, the equivalence between the two systems is justified by a rigorous theoretical concept. 
This similarity concept is less constraining since additional free parameters are introduced. Nevertheless it better corresponds to the real problematic of laboratory astrophysics rescaling. The use of this similarity concepts claims a physical justification that the unconserved sub-physical scales do not modify the dynamics of plasmas. This last approach opens fundamental perspectives since phenomena which cannot be reproduced according to the \textit{absolute similarity} concepts become reproducible \citep{Falize10}. These theoretical considerations are applied to the dynamics of different high-energy density radiating regimes. 

\section{Scalability properties of optically thin radiating plasmas}

A plasma can have density and high-temperature conditions so that an important part of energy is radiated in the form of low-interacting radiation ($\tau<< 1$ where $\tau$ is the optical depth). The optically thin regime concerns a great variety of astrophysical phenomena, especially the observables ones, such as  the  first stage of molecular contraction,
the dynamics of stellar jets and outflows, radiative accretion shocks, the late supernovae remnants and galaxy formation. 
The radiative cooling can greatly modify the structure, the dynamics and the stability of the emitting plasmas. According to the properties of cooling processes, the cooling instability can develop \citep{Lyndenbell00}. It attracts many astrophysicists attention, since it explains the clumpy structure of interstellar medium (ISM) and the co-existence of the cold neutral medium and warm diffuse medium phases in ISM.\\
Different experimental studies of radiative jet collapse \citep{Shigemori00, Gregory10b} and the cooling instability \citep{Moore05} have been realized with powerful lasers. The observation of dense localized structure of lower temperature formed by this instability is very common in Z-pinch and tokamak experiments \citep{Meerson96}.  
Given the various astrophysical environments concerned by this regime and the experimental possibilities producing equivalent plasmas, the study of their scalability is essential. 
A simple modeling can be done by introducing a loss of entropy and the dynamics of plasma is given by the following equations:

\begin{equation}\label{eq3_1:eq}
\frac{\partial \rho}{\partial t}+\vec{\nabla}.(\rho \vec{v})=0,\quad dM=\rho.dV \; ,
\end{equation}

\begin{equation}\label{eq3_2:eq}
 \rho\frac{d\vec{v}}{dt}=-\vec{\nabla}P, \quad \frac{d}{dt}=\left[\frac{\partial }{\partial t}+(\vec{v}.\vec{\nabla})\right] \; ,
\end{equation}

\begin{equation}\label{eq3_3:eq}
 \frac{d P}{dt}-\gamma\frac{P}{\rho}\frac{d\rho}{dt}=-(\gamma-1)\mathcal{L}(\rho,T) \; ,
\end{equation}
where $t$, $\vec{v}$, $M$, $V$, $\rho$, $P$, $\gamma$ and $\mathcal{L}(\rho,T)$ are respectively time, velocity, mass, volume, density, thermal pressure, adiabatic index of plasma and cooling function. In this paper, the function $\mathcal{L}(\rho,T)$ is chosen as $\mathcal{L}(\rho,T) = \mathcal{Q}_{1}(\rho,T)+\mathcal{Q}_{2}(\rho,T)$
where $\mathcal{Q}_{1}$ and $\mathcal{Q}_{2}$ are energy sources (or losses) in order to take into account two different radiating physical processes.
The source terms are supposed to take an analytical form given by:
\begin{equation} \label{eq3_Q}
\mathcal{Q}_{i}(\rho,P)=\mathcal{Q}_{0,i}\rho^{\epsilon_{i}}P^{\zeta_{i}}r^{\theta_{i}},
\end{equation}
where  $r$, $\mathcal{Q}_{0,i}$, $\epsilon_{i}$, $\zeta_{i}$ and $\theta_{i}$ are respectively the spatial coordinate and four characteristic constants of source processes. This form generalizes the optically thin case where $\theta_{i}=0$. 
Thanks to the spatial dependence of Eq.~(\ref{eq3_Q}), we can approximatively model optically thick processes  \citep{Chanmugam85}.
The analytical form of $\mathcal{Q}_{i}$  in Eq.~(\ref{eq3_Q}) is motivated by the first fact that several continuous processes can be modeled exactly or approximatively by power laws and by the second fact that $\mathcal{Q}_{i}\propto \kappa_{P}\sigma T^{4}$ which can be modeled by a power law at high-temperature (where $\sigma$ is the Stefan-Boltzmann constant and $\kappa_{P}$ is the Planck opacity).
Although the ISM cooling function takes a very complex form \citep{Dalgarno72}, it can be approximated by a power law model in several temperature regimes.\\ 
In order to write the energy evolution in Eq.~(\ref{eq3_3:eq}), a polytropic evolution of plasma have been assumed:
\begin{equation}\label{eq3_4:eq}
\rho e=\frac{P}{\gamma-1},
\end{equation}
  where $e$ is the specific internal energy. In order to close the equation system, an EOS should be added. The pressure relation (\ref{eq3_4:eq}) holds for a larger class of EOS and not only for an ideal gas \citep{Ryutov01}.   
In this paper, we consider an EOS given by \citet{Zeldo}:
\begin{equation}\label{eq:eos}
P=\varepsilon_{0}(Z)\rho^{\mu}T^{\nu} \; , 
\end{equation}
where $\varepsilon_{0}(Z)$, $\mu$, $\nu$ are respectively a function of the ionization $Z$ and two exponents to be chosen later on. \citet{Zeldo} have noticed that to keep the consistency of the thermodynamic description of gases, we should have:
\begin{equation}\label{adiabatic:eq}
\gamma=\frac{\nu-\mu}{\nu-1} \; .
\end{equation}
We easily verify that a photon gas, which is characterized by a pressure $P_{r}=a_{r} T^{4}/3$, where $a_{r}$ is the radiative constant, verify this constraint.\\ 
In order to establish the generic scaling laws, the relation between the typical quantities in astrophysical objects and laboratory experiments are given by:  
\begin{equation}\label{eq:redim1}
r=\lambda^{\delta_{1}}\tilde{r},\quad t=\lambda^{\delta_{2}}\tilde{t}, \quad \vec{v}=\lambda^{\delta_{3}}\tilde{\vec{v}}, \quad M=\lambda^{\delta_{4}}\tilde{M},
\end{equation}

\begin{equation}
\rho=\lambda^{\delta_{5}}\tilde{\rho},\quad P=\lambda^{\delta_{6}}\tilde{P},\quad T=\lambda^{\delta_{7}}\tilde{T},\quad \gamma=\lambda^{\delta_{8}}\tilde{\gamma} \; ,
\end{equation}

\begin{equation}\label{eq:redim3}
\varepsilon_{0}=\lambda^{\delta_{9}}\tilde{\varepsilon}_{0} , \quad \mathcal{Q}_{0,1}=\lambda^{\delta_{10}}\tilde{Q}_{0,1}, \quad \mathcal{Q}_{0,2}=\lambda^{\delta_{11}}\tilde{Q}_{0,2} \; .
\end{equation}
Not only the adiabatic index must be invariant, but also the classical hydrodynamic dimensionless numbers must be preserved:
\begin{equation}\label{eq1:eq}
St=\frac{v t}{r}, \quad \mathcal{M}=\frac{v}{c_{s}},
\end{equation}
where $St$ and $\mathcal{M}$ are respectively the Strouhal and Mach number. Moreover, a radiation dimensionless number, the cooling (or heating) parameter $\chi_{\mathcal{L}}$, must be invariant in order to conserve the balance between the radiation and hydrodynamic effects. It is defined by:
\begin{equation}
\chi_{\mathcal{L}}=\frac{t_{\mathcal{L}}}{t}=\frac{P}{(\gamma-1) \mathcal{L} t}\;, 
\end{equation}
which leads, for the cooling function considered in this section, to the following results:
\begin{equation}\label{eq_1:eq}
\chi_{\mathcal{Q}_{1}}=\frac{P}{(\gamma-1) \mathcal{Q}_{1} t}\;, \quad \chi_{\mathcal{Q}_{2}}=\frac{P}{(\gamma-1) \mathcal{Q}_{2} t}\;, 
\end{equation}
Although the Strouhal number is meaningful only when two flows are being  compared, the others give informations about the studied plasma itself. Indeed, if $\mathcal{M}>1$ (or $\mathcal{M}<1$), the flow is supersonic (or subsonic) and if $\chi_{\mathcal{L}}<1$ (or $\chi_{\mathcal{L}}>1$) the flow is radiating (or adiabatic). One important result from the similarity study of these radiating plasmas  is the conservation of the exponents of the cooling function. 
Since the cooling instability criterion \citep{Lyndenbell00} or the complex dynamics of these fluids greatly depends on the exponents of the cooling function, their conservation is very important in the context of laboratory studies.\\
Introducing Eqs.~(\ref{eq:redim1}-\ref{eq:redim3}) in Eqs.~(\ref{eq3_1:eq}-\ref{eq3_3:eq}), we have obtained analytically the scaling laws insuring the invariance of Eqs.~(\ref{eq3_1:eq}-\ref{eq3_3:eq}). In Table \ref{Table1:tab} different general scaling laws are presented. In the second column we present the scaling laws obtained with \textit{global similarity} concept for a composite generalized cooling function, although in the third column the scaling laws of purely optically thin radiating plasmas are presented. In Table \ref{Table2:tab} scaling laws of different astrophysical systems are presented in \textit{absolute similarity} case. Although two free parameters (noted $\delta_{5}$, $\delta_{6}$) are obtained when the scaling laws are constructed from the \textit{absolute similarity} concepts ($\mathcal{Q}_{0,i}$ and $\varepsilon_{0}$ are invariants), four free parameters ($\delta_{1}$, $\delta_{5}$, $\delta_{6}$, $\delta_{9}$) are obtained when the \textit{global similarity} concept is used as in the hydrodynamic case \citep{Falize09b}.\\
Here, the results are given for two astrophysical cases: supernova remnants and accretion shock in magnetic cataclysmic variables.\\
The scalability properties of supernova remnants in radiative phase are presented in the second column. In the temperature regime of remnants, the cooling function is approximatively given by $\mathcal{L}\propto \rho^{2}T^{-1/2}$ which is the expression used in this application.\\
The scalability properties of accretion column in magnetic cataclysmic variables are presented in the third column (with bremsstrahlung emission) and the fourth column (with cyclotron and bremsstrahlung emissions). In these astrophysical objects the X-ray emitting regions are located near the magnetic poles, where the matter is heated by a stand-off shock to a temperature of around 10-50 keV, then is cooled by bremsstrahlung emission ($\mathcal{Q}\propto \rho^{2}T^{1/2}$) and other cooling processes (as cyclotronic emission \citep{Saxton99} $\mathcal{Q}\propto \rho^{0.15}T^{2.5}$). These radiation losses lead to the formation of a cooling layer \citep{Chevalier82}. In this complex zone, named the accretion column, the presence of an intense magnetic field, radiation, and hydrodynamics leads to a rich range of behaviors at different spatial and temporal scales. The accretion column presents a highly stratified structure in temperature and density, which depends greatly on the physical properties of the white dwarf \citep{Wu95}.
Unfortunately, the size scales associated with these zones are of the order of the white dwarf radius or smaller, which complicates their direct observation \citep{Hoogerwerf06}. These high-energy environments present interesting scalability properties since the main radiating processes can be modeled by a power law form. Noting that, in some AM Her stars (mCVs with $B>10$ MG), the accretion column is dominated by bremsstrahlung cooling, implying that the magnetic field acts only to guide the plasma and does not modify the local dynamics. Thus, by the results of Table \ref{Table2:tab},  we demonstrate that an adapted scaling law allows to produce, with powerful lasers, a diagnosable accreting column in the laboratory, and to study its structure. This is also justified by the fact that the accretion column height, $L_{h}$, is given by $L_{h}\sim v_{s}\times t_{cool}$ with $v_{s}$ and $t_{cool}$ as the velocity of accretion matter and the cooling time. For typical laboratory regime, $v_{s}\sim 100$ km.s$^{-1}$ and $t_{cool}\sim 1$ ns, the height of the accretion shock is $100$ $\mu$m which is a diagnosable scale. This has recently been investigated experimentally with the LULI2000 facility \citep{Falize10a} which constitutes the first laser experiment that aims at producing relevant accreted columns in laboratory. 
 If such a goal is reached, the knowledge gained in the laboratory can be applied to similar astrophysical processes.

\section{Scalability properties of optically thick radiating plasmas}
Several  high-energy astrophysics phenomena present a highly coupled physical regime between radiation and matter which leads to a complex structure and dynamics of the flow. 
It is the case for accretion discs, stellar interiors, supernova shocks and also the evaporation of clouds in ISM.
In spite of the access to multi-wavelengths information, the understanding of these objects is generally partial. With adapted target designs and compositions, 
relevant conditions of some phenomena are, nowadays, commonly created and diagnosed with the modern powerful facilities.  Indeed the ability to produce intense X-ray radiation and to diagnose its interaction with matter, or to create strong shocks which lead to intense emitted radiation, constitute a real opportunity  to test and validate the physical models of such structure ubiquitous in astrophysical environments. In this section two specific radiation hydrodynamic regimes are studied in detail. We examine the scalability properties of optically thick radiating plasmas in one-temperature diffusive regimes. We firstly focus on the regime where the energy transport is efficiently performed by the radiation and secondly the regime where the radiation field is so high that the radiative energy density and pressure are not negligible to counterpart matter. The diffusion approximation is correct when the radiative Knudsen number, $ Kn_{r}$,  is small, which is defined by: 
 \begin{equation}
 Kn_{r}=\frac{l_{R}(\rho,T)}{L_{H}},
 \end{equation}
 where $l_{R}$, $L_{H}$ are respectively the mean free path of radiation and the hydrodynamical scale. In order to insure that the radiation and matter are at the same temperature, the mean free path should be smaller than the typical temperature gradient length $l_{T}$:
 \begin{equation}
 \frac{l_{R}}{l_{T}}=\frac{l_{R}(\rho,T)}{T}\nabla T<<1.
 \end{equation}
Determining the radiating regime of fluids, two dimensionless numbers are commonly introduced. 
In order to evaluate the efficiency of radiation, the enthalpy flux, $\rho h v$, is compared to the black body radiative flux, $\sigma T^{4}$ in the conventional Boltzmann number \citep{Mihalas99, Castor04}:
\begin{equation}
Bo=\frac{\rho h v}{\sigma T^{4}}.
\end{equation}
Thus the energy is transported efficiently by radiation when the temperature of plasma is greater than a critical temperature, $T_{Bo}$, corresponding to the case $Bo=1$:
\begin{equation}\label{eq54:eq}
T\geqslant T_{Bo}\equiv\left[\frac{\gamma}{\gamma-1}\right]^{1/3}\left[\frac{k_{B}}{\sigma \mu m_{H}}\right]^{1/3}\left[\rho v\right]^{1/3},
\end{equation}
where $k_{B}$ and $m_{H}$ are respectively the Boltzmann constant and the mass of the hydrogen atom.
The expression (\ref{eq54:eq}) gives for a perfect gas with an interstellar composition in conventional units:
\begin{equation}\label{eq55:eq}
T_{Bo}[\textrm{keV}]= 5.18\times 10^{-2}\left[\frac{\rho}{1 \textrm{g.cm}^{-3}}\right]^{1/3}\left[\frac{v}{1 \textrm{km.s}^{-1}}\right]^{1/3}.
\end{equation}
For the common radiative regime obtained in laboratory with kJ facilities \citep{Koenig06, Michaut07, Michaut09}, the critical temperature is around 40 eV. 
When the characteristic velocity is the sound velocity, the relation (\ref{eq54:eq}) is written only in function of the density of material by:  

\begin{equation}
T_{Bo}= \left[\frac{\gamma\sigma^{-1}\sqrt{\gamma}}{\gamma-1}\right]^{2/5}\left[\frac{k_{B}}{\mu m_{H}}\right]^{3/5}\rho^{2/5}.
\end{equation}
For a polytropic ideal gas ($\gamma=5/3$) with an interstellar composition the critical temperature is given by:
\begin{equation}
T_{Bo}[\textrm{keV}]= 5.86\times10^{-1}\left[\frac{\rho}{1 \textrm{g.cm}^{-3}}\right]^{2/5}.
\end{equation}
Another important dimensionless number is the so-called Mihalas number, $R$, which is defined as the ratio of the material internal energy density ($\rho e$) to the radiation energy density ($E_{r}$): 
\begin{equation}\label{mihalas:eq}
R=\frac{\rho e}{E_{r}}=\frac{1}{\gamma-1}\frac{P}{a_{R}T^{4}}.
\end{equation}
 It measures the relative importance of gas and radiation pressure since $E_{r}$ is proportional to the radiative pressure by the Eddington approximation. As for the Boltzmann number, a critical temperature can be determined from it. Thus the radiation plays an important role in laboratory when the temperature satisfies the following criterion: 
\begin{equation}\label{temp-mihalas:eq}
T\geqslant T_{R}\equiv\left[\frac{k_{B}}{\mu m_{H} a_{R}(\gamma-1)}\right]^{1/3}\rho^{1/3},
\end{equation}
\begin{equation}\label{eq_tem_m:eq}
T_{R}[\textrm{keV}] = 3.88 \left[\frac{\rho}{1 \textrm{g.cm}^{-3}}\right]^{1/3}.
\end{equation}
Using typical values of foam target ($\rho\sim 0.1$ g.cm$^{-3}$), $T_{R}$ is around 1 keV. For a gas target the critical temperature is lower and more easily achieved. 
Such extreme radiating regimes will be commonly created on NIF or LMJ facilities. Consequently it is very important to study the similarity properties of such radiating fluids. 

\subsection{The radiative flux regime}
We begin by examining the scalability properties of radiating plasmas in radiative flux regime ($Bo<1$, $R>1$). 
 In this case Eq.~(\ref{eq3_3:eq}) changes and takes the following form:
\begin{equation}\label{eq4_1:eq}
 \frac{dP}{dt}-\gamma\frac{P}{\rho}\frac{d\rho}{dt}=-(\gamma-1)\vec{\nabla}.\vec{F}_{r}-(\gamma-1)\mathcal{Q} \; ,
\end{equation}
where $\vec{F}_{r}$ is the radiative flux and $\mathcal{Q}=\mathcal{Q}_{0}\rho^{\epsilon}P^{\zeta}r^{\theta}$ is similar to the quantity arising in the previous section. Taking a general form allows to include another kind of radiative flux term provided the condition $\theta=-2$ holds. 
In the diffusion regime the radiative flux is given by: 
\begin{equation}\label{eq4_2:eq}
\vec{F}_{r}=-\frac{l_{R}(\rho,T)c}{3} \, \vec{\nabla} E_{r}=-\kappa_{r}(\rho,T) \, \vec{\nabla} T \; ,
\end{equation}
where $c$ and $\kappa_{r}$ are respectively the light celerity and the radiative conductibility. We assume that $\kappa_{r}(\rho,T)$ must be reduced, in the thermodynamical regime of interest here, to a power law form:
\begin{equation}
\kappa_{r}(\rho,T)=\kappa_{0}\rho^{m}T^{n},
\end{equation}
where $\kappa_{0}$, $m$ and $n$ are three constant coefficients characterizing the radiative process.
This form is motivated by the scalability properties but also, as in the cooling function case, because several radiative processes can be modeled approximatively or exactly by such a form. 
   The scalability properties of such flows can be constructed using the transformation (\ref{eq:redim1}-\ref{eq:redim3}) with the additional relation
$ \kappa_{0}=\lambda^{\delta_{12}}\tilde{\kappa}_{0}$.
  From the similarity properties, a new dimensionless number, $\Pi$,  is added to the previous numbers, which writes:
\begin{equation}\label{eq_bo:eq}
\Pi=\frac{P}{{F}_{r}}\frac{x}{t}=\frac{3}{16}\frac{\gamma-1}{\gamma}\frac{l_{T}}{l_{R}}\frac{Bo}{St}. \quad
\end{equation}
Actually, the quantity $l_{T}Bo/l_{R}$ must be invariant but if $Bo$ is conserved, the ratio $l_{T}/l_{R}$ must be an invariant too ($l_{T}/l_{R}=\tilde{l}_{T}/\tilde{l}_{R}$).
As previously, the scaling laws have been calculated from invariance properties of Eqs.~(\ref{eq3_1:eq}, \ref{eq3_2:eq}, \ref{eq4_1:eq}) and are presented in Table \ref{Table3:tab}. It is straightforward to show that the equations are invariant under the following scale transformation. In other words, the equation forms describing the dynamics of the astrophysical system and the laboratory plasmas  are indistinguishable through the scale transformation.\\
In the second column of Table \ref{Table3:tab}, the scaling laws in the \textit{global similarity} case are presented.  As in previous radiating regimes, four free parameters ($\delta_{1}$, $\delta_{5}$, $\delta_{6}$, $\delta_{9}$) are obtained in order to scale an experiment. 
In the purely radiative flux regime ($\mathcal{Q}=0$) and for the \textit{absolute similarity}, the radiative flux imposes a complementary constraint leading to a reduced number of free parameters.  Two free parameters ($\delta_{5}$, $\delta_{6}$) are obtained and corresponding scaling laws are presented in the third column.\\
In Table \ref{Table4:tab}, we provide the scaling laws when the radiative transport is respectively modeled by Spitzer conduction ($\kappa_{r}\propto T^{5/2}$), Bridgman limit of thermal conduction ($\kappa_{r}\propto \rho^{2/3}T^{1/2}$), Dyson radiative limit ($\kappa_{r}\propto \rho^{-1} T^{4}$), Thomson scattering ($\kappa_{r}\propto \rho^{-1} T^{3}$), Kramers opacity ($\kappa_{r}\propto \rho^{-2} T^{13/2}$) and dust grains ($\kappa_{r}\propto \rho^{-1}T$). 
Thus, this analysis shows the attractive perspectives of laboratory experiments in order to reproduce astrophysical phenomena in this specific radiating regime.  
Various scaling laws are derived by \citet{Murakami02} in the context of inertial confinement fusion for a internal energy relation in the form: $e\propto T^{\beta}$ where $\beta$ is an arbitrary exponent. The fundamental problem of such EOS is that it does not preserve the thermodynamic consistency of the gas \citep{Zeldo} contrary to EOS used in this paper. 
Finally, noting that the Rankine-Hugoniot relations for hybrid radiative shock \citep{Michaut09} are invariant by the general scale transformations presented in Table \ref{Table3:tab}, we theoretically prove that they can be reproduced in laboratory experiments.

\subsection{The  fully radiative regime}
Now the scalability properties of radiating fluids, when the radiative energy density and pressure are important compared to their matter counterpart, are examined.  This regime concerns the explosion phase of supernovae, several accretion flows, fundamental phase in star formation, in stellar mass losses or the ablation of molecular clouds \citep{Konigl84} in ISM. In massive stars the radiation quantities  become of the same order of magnitude as the thermal ones when the mass of star is around $30\,M_{\odot}$ \citep{Chandrasekhar03}. 
In addition to mass conservation (see Eq.~(\ref{eq3_1:eq})), the plasma evolution is governed by the following equations
 \citep{Pomraning73, Coggeshall86, Drake06}:
\begin{equation}\label{eq4_3:eq}
\rho\frac{d\vec{v}}{dt}=-\vec{\nabla}P_{T} \; , 
\end{equation}
\begin{equation}\label{eq4_4:eq}
\quad \frac{dE_{T}}{dt}-\frac{E_{T}+P_{T}}{\rho}\frac{d\rho}{dt}=-\vec{\nabla}.\vec{F}_{r}-\mathcal{Q} \; ,
\end{equation}
where $E_{T}$ and $P_{T}$ are respectively the total energy density and pressure, given by:
\begin{equation}\label{eq4_5:eq}
E_{T}=\rho e+E_{r}, \quad P_{T}=P+P_{r}.
\end{equation}
In the present application $E_{r}=a_{R}T^{4}$ and $P_{r}=E_{r}/3$.\\
By the similarity properties the main characteristic dimensionless numbers are identified. The Mihalas number given by Eq.~(\ref{mihalas:eq}) is added to the four previous dimensionless numbers Eqs.~(\ref{eq1:eq}), (\ref{eq_1:eq}) and (\ref{eq_bo:eq}). The corresponding scaling laws are presented in Table~\ref{Table5:tab}. The second column corresponds to the \textit{global similarity} case where three free parameters ($\delta_{5}$, $\delta_{9}$, $\delta_{12}$) are found. The loss of one free parameter, compared to the previous \textit{global similarity} cases, comes from the fact that $P_{r}$ and $E_{r}$ introduce a new fundamental constant which is not scalable. In the \textit{absolute similarity} case one homothetic group is found with one free parameter ($\delta_{5}$) which can be chosen arbitrarily. This latter defines the magnitude of the other characteristic physical quantities which has to be maintained in order to insure that the scale plasma behaves similarly to the astrophysical phenomena. In Table \ref{Table5:tab} we have chosen to write all the quantities in terms of the ratio of density in order to obtain a simple generic expression. It is the first time that the possibility  of reproducing an exact scale model of astrophysical phenomena in such regime is demonstrated. This is an important result since it opens new and important opportunities for laboratory astrophysics experiments for studying the dynamics of plasmas in such regimes. In the fourth and fifth columns, scale transformations are proposed in two important cases with dust and Kramers opacity. 
Since the Rankine-Hugoniot relations \citep{Mihalas99,Bouquet00} are necessarily scale invariant, the existence of scaling laws allows to demonstrate that radiative shocks in fully radiative regime \citep{Michaut09} can be theoretically reproduced in laboratory experiments. 
\section{Conclusion}
This paper presents the scalability properties of radiation hydrodynamic fluids and proposes new scaling laws. In this work, an exhaustive description of similarity concepts is presented and new invariance concepts are introduced remaining more or less the physics at sub-scales. It is important to master the subtleties of the \textit{absolute similarity} and the \textit{global similarity}.
Currently, the constraints imposed by the \textit{absolute similarity}, which is more rigorous, are very restrictive for astrophysical laboratory applications due to great number of constraints. Consequently, the \textit{global similarity} is preferred and is an important theoretical support to design an astrophysical experiment. In spite of the \textit{absolute similarity} is very interesting in several high-energy density applications for adapting the target design in more and more powerful facilities. \\
This work constitutes a fundamental and powerful tool determining the astrophysical relevance of modern high-energy density laboratory experiments. We have examined three types of radiative regimes: the optically thin regime, the optically thick one in which the radiative flux regime is distinguished from the fully radiative one. The possibility of reproducing a scaled model of radiating plasmas with a low Mihalas number, \textit{i.e.} the fully radiative regime, is a real opportunity to progress in the understanding of the induced complex physics. For the first time, the scaling laws are rigorously demonstrated for such flows occurring in several extreme astrophysical environments. More generally, we have showed that a broad class of astrophysical radiating plasmas for optically thin regime as well as the two specific optically thick regimes can be simulated in high-energy density laboratory experiments.\\
The key results presented here prove that laboratory astrophysics is a very promising and fruitful approach that can improve, complete and test our understanding of physical mechanisms acting in high-energy astrophysical environments.

\newpage
\appendix

\clearpage

\clearpage

\begin{deluxetable}{llllllllllllcrl}
\tabletypesize{\scriptsize}
\tablecaption{\emph{Scaling laws of optically thin radiating fluids. The scaling laws of generalized cooling function case obtained using the \textit{global similarity} are presented in the second column. The scaling of purely optically thin plasmas are showed in the third column.}}
\tablewidth{0pt}
\tablehead{
 \colhead{physical ratio} & \colhead{ global similarity case} & global similarity with $\theta_{i}=0$ 
}
\startdata
$r/\tilde{r}$ & $\lambda^{\delta_{1}}$ & $\lambda^{\delta_{1}}$  \\
$t/\tilde{t}$& $\lambda^{\delta_{1}+(\delta_{5}-\delta_{6})/2}$ &   $\lambda^{\delta_{1}+(\delta_{5}-\delta_{6})/2}$  \\
$v/\tilde{v}$ & $\lambda^{(\delta_{6}-\delta_{5})/2}$  &  $\lambda^{(\delta_{6}-\delta_{5})/2}$   \\
$\rho/\tilde{\rho}$ & $\lambda^{\delta_{5}}$  &$\lambda^{\delta_{5}}$ \\
$P/\tilde{P}$ & $\lambda^{\delta_{6}}$  &  $\lambda^{\delta_{6}}$    \\
$T/\tilde{T}$ & $\lambda^{(\delta_{6}-\delta_{9}-\mu \delta_{5})/\nu}$ &  $\lambda^{(\delta_{6}-\delta_{9}-\mu \delta_{5})/\nu}$   \\
$\mathcal{Q}_{0,1}/\tilde{\mathcal{Q}}_{0,1}$ &  $\lambda^{(3/2-\zeta_{1})\delta_{6}-(\epsilon_{1}+1/2)\delta_{5}-(\theta_{1}+1)\delta_{1}}$ & $\lambda^{(3/2-\zeta_{1})\delta_{6}-(\epsilon_{1}+1/2)\delta_{5}-\delta_{1}}$  \\
$\mathcal{Q}_{0,2}/\tilde{\mathcal{Q}}_{0,2}$  &  $\lambda^{(3/2-\zeta_{2})\delta_{6}-(\epsilon_{2}+1/2)\delta_{5}-(\theta_{2}+1)\delta_{1}}$ & $\lambda^{(3/2-\zeta_{2})\delta_{6}-(\epsilon_{2}+1/2)\delta_{5}-\delta_{1}}$   \\
$\varepsilon_{0}/\tilde{\varepsilon}_{0}$ & $\lambda^{\delta_{9}}$ & $\lambda^{\delta_{9}}$ \\
\label{Table1:tab}
\enddata
\end{deluxetable}

\begin{deluxetable}{llllllllllllcrl}
\tabletypesize{\scriptsize}
\tablecaption{\emph{Scaling laws of optically thin plasmas for different astrophysical applications are presented. The scaling laws of radiative supernova remnant, accretion shock with bremsstrahlung cooling and accretion with bremsstrahlung and cyclotron cooling  are respectively shown in the second, third and fourth columns.}}
\tablewidth{0pt}
\tablehead{
 \colhead{physical ratio} & \colhead{radiative SNR regime}  & \colhead{BC} & \colhead{BC+CC} 
}
\startdata
$r/\tilde{r}$ & $\lambda^{2\delta_{6}-3\delta_{5}}$  & $\lambda^{\delta_{6}-2\delta_{5}}$  & $\lambda^{-\frac{3}{40}\delta_{5}}$ \\
$t/\tilde{t}$&  $\lambda^{\frac{3}{2}\delta_{6}-\frac{5}{2}\delta_{5}}$ & $\lambda^{\frac{1}{2}\delta_{6}-\frac{3}{2}\delta_{5}}$  &  $\lambda^{-\frac{43}{80}\delta_{5}}$ \\
$v/\tilde{v}$ & $\lambda^{\frac{1}{2}\delta_{6}-\frac{1}{2}\delta_{5}}$  & $\lambda^{\frac{1}{2}\delta_{6}-\frac{1}{2}\delta_{5}}$  & $\lambda^{\frac{37}{80}\delta_{5}}$  \\
$\rho/\tilde{\rho}$ &  $\lambda^{\delta_{5}}$  & $\lambda^{\delta_{5}}$  &$\lambda^{\delta_{5}}$ \\
$P/\tilde{P}$ & $\lambda^{\delta_{6}}$ &$\lambda^{\delta_{6}}$ &$\lambda^{\frac{77}{40}\delta_{5}}$   \\
$T/\tilde{T}$ & $\lambda^{\delta_{6}-\delta_{5}}$ & $\lambda^{\delta_{6}-\delta_{5}}$  &$\lambda^{\frac{37}{40}\delta_{5}}$  \\
$\mathcal{Q}_{0,1}/\tilde{\mathcal{Q}}_{0,1}$ &  1 & 1   & 1 \\
$\mathcal{Q}_{0,2}/\tilde{\mathcal{Q}}_{0,2}$  & --- &  ---  & 1  \\
$\varepsilon_{0}/\tilde{\varepsilon}_{0}$  & 1  & 1 & 1\\
\label{Table2:tab}
\enddata
\end{deluxetable}

\begin{deluxetable}{llllllllllllcrl}
\tabletypesize{\scriptsize}
\tablecaption{\emph{General scaling laws for radiative flux regime. The scaling laws obtained using the \textit{global similarity} and the \textit{absolute similarity} are respectively presented in the third and fourth columns.} }
\tablewidth{0pt}
\tablehead{
 \colhead{physical ratio} & \colhead{global similarity case}  & \colhead{Purely radiative flux regime}
}
\startdata
$r/\tilde{r}$ & $\lambda^{\delta_{1}}$ & $\lambda^{[m+1/2-(n+1)\mu/\nu]\delta_{5}+[(n+1)/\nu-3/2]\delta_{6}}$  \\
$t/\tilde{t}$ & $\lambda^{\delta_{1}+(\delta_{5}-\delta_{6})/2}$ & $\lambda^{[m+1-(n+1)\mu/\nu]\delta_{5}+[(n+1)/\nu-2]\delta_{6}}$   \\
$\rho/\tilde{\rho}$ & $\lambda^{\delta_{5}}$  & $\lambda^{\delta_{5}}$  \\
$v/\tilde{v}$ & $\lambda^{(\delta_{6}-\delta_{5})/2}$ & $\lambda^{(\delta_{6}-\delta_{5})/2}$  \\
$P/\tilde{P}$ & $\lambda^{\delta_{6}}$ & $\lambda^{\delta_{6}}$\\
$T/\tilde{T}$  & $\lambda^{(\delta_{6}-\mu\delta_{5}-\delta_{9})/\nu}$ & $\lambda^{(\delta_{6}-\mu\delta_{5})/\nu}$  \\
$\varepsilon_{0}/\tilde{\varepsilon}_{0}$ & $\lambda^{\delta_{9}}$ & 1 \\
$\mathcal{Q}_{0}/\tilde{\mathcal{Q}}_{0}$ & $\lambda^{(3/2-\zeta)\delta_{6}-(\epsilon+1/2)\delta_{5}-(\theta+1)\delta_{1}}$ & --- \\
$\kappa_{0}/\tilde{\kappa}_{0}$ &  $\lambda^{\delta_{1}+[(n+1)/\nu]\delta_{9}+[3/2-1/\nu-n/\nu]\delta_{6}+[\mu/\nu-1/2-m+n\mu/\nu]\delta_{5}}$ & 1 \\
\label{Table3:tab}
\enddata
\end{deluxetable}

\begin{deluxetable}{llllllllllllcrl}
\tabletypesize{\scriptsize}
\tablecaption{\emph{Scaling laws of various optically thick radiating fluids. The Spitzer, Bridgman, Kramers,  Dyson, Thomson and Dust grain cases are respectively presented in the second, third, fourth, fifth, sixth and seventh columns.}}
\tablewidth{0pt}
\tablehead{
 \colhead{physical ratio} & \colhead{Spitzer} & \colhead{Bridgman} & \colhead{Kramers} & \colhead{Dyson} & \colhead{Thomson} & \colhead{Dust grain}
}
\startdata
$r/\tilde{r}$  & $\lambda^{2\delta_{6}-3\delta_{5}}$ & $\lambda^{-\delta_{5}/3}$  & $\lambda^{-9\delta_{5}+6\delta_{6}}$ & $\lambda^{7\delta_{6}/2-11\delta_{5}/2}$ & $\lambda^{5\delta_{6}/2-9\delta_{5}/2}$ & $\lambda^{\delta_{6}/2-5\delta_{5}/2}$ \\
$t/\tilde{t}$ & $\lambda^{3\delta_{6}/2-5\delta_{5}/2}$  & $\lambda^{\delta_{5}/6-\delta_{6}/2}$ & $\lambda^{-17\delta_{5}/2+11\delta_{6}/2}$ & $\lambda^{3\delta_{6}-5\delta_{5}}$ & $\lambda^{2\delta_{6}-4\delta_{5}}$ & $\lambda^{-2\delta_{5}}$ \\
$v/\tilde{v}$ & $\lambda^{\delta_{6}/2-\delta_{5}/2}$ & $\lambda^{\delta_{6}/2-\delta_{5}/2}$ & $\lambda^{\delta_{6}/2-\delta_{5}/2}$ & $\lambda^{\delta_{6}/2-\delta_{5}/2}$ & $\lambda^{\delta_{6}/2-\delta_{5}/2}$ & $\lambda^{\delta_{6}/2-\delta_{5}/2}$  \\
$\rho/\tilde{\rho}$ &  $\lambda^{\delta_{5}}$ & $\lambda^{\delta_{5}}$ & $\lambda^{\delta_{5}}$ & $\lambda^{\delta_{5}}$ & $\lambda^{\delta_{5}}$ & $\lambda^{\delta_{5}}$ \\
$P/\tilde{P}$  & $\lambda^{\delta_{6}}$  & $\lambda^{\delta_{6}}$ & $\lambda^{\delta_{6}}$& $\lambda^{\delta_{6}}$ & $\lambda^{\delta_{6}}$ & $\lambda^{\delta_{6}}$ \\
$T/\tilde{T}$   & $\lambda^{\delta_{6}-\delta_{5}}$ & $\lambda^{\delta_{6}-\delta_{5}}$ & $\lambda^{\delta_{6}-\delta_{5}}$ & $\lambda^{\delta_{6}-\delta_{5}}$ & $\lambda^{\delta_{6}-\delta_{5}}$ & $\lambda^{\delta_{6}-\delta_{5}}$ \\
$\mathcal{Q}_{0}/\tilde{\mathcal{Q}}_{0}$  & --- & --- & --- & --- & --- & ---  \\
$\varepsilon_{0}/\tilde{\varepsilon}_{0}$ & $1$ & $1$ & $1$  & $1$ & $1$ & $1$  \\
$\kappa_{0}/\tilde{\kappa}_{0}$  & $1$ & $1$ & $1$ & $1$ & $1$ & $1$  \\
\label{Table4:tab}
\enddata
\end{deluxetable}

\begin{deluxetable}{lllllllllllllrl}
\tabletypesize{\scriptsize}
\tablecaption{\emph{Scaling laws for optically thick radiating fluids in the fully radiative regime. The scaling laws obtained using the global similarity are presented in the second column. In the third column, the scaling laws of Zeldovich -Raizer gas obtained using the absolute similarity are showed. In the fourth and fifth columns, two applications for ideal gas (Ig) are presented.  }}
\tablewidth{0pt}
\tablehead{
 \colhead{physical ratio} & \colhead{global similarity case} & \colhead{Zeldovich-Raizer gas} & \colhead{Ig + dust} & \colhead{Ig + Kramers op.}
}
\startdata
$r/\tilde{r}$ & $\lambda^{\delta_{12}+([n-5]/[4-\nu])\delta_{9}+([m+1/2]+\mu[n-5]/[4-\nu])\delta_{5}}$ & $\lambda^{([m+1/2]+\mu[n-5]/[4-\nu])\delta_{5}}$  & $\lambda^{-11\delta_{5}/6}$ & $\lambda^{-\delta_{5}}$ \\
$t/\tilde{t}$& $\lambda^{\delta_{12}+([n-7]/[4-\nu])\delta_{9}+(m+1+\mu[n-7]/[4-\nu])\delta_{5}}$  &  $\lambda^{(m+1+\mu[n-7]/[4-\nu])\delta_{5}}$ & $\lambda^{-2\delta_{5}}$ & $\lambda^{-7\delta_{5}/6}$ \\
$v/\tilde{v}$ & $\lambda^{(2/[4-\nu])\delta_{9}+([4\mu+\nu-4]/[8-2\nu])\delta_{5}}$   & $\lambda^{([4\mu+\nu-4]/[8-2\nu])\delta_{5}}$ & $\lambda^{\delta_{5}/6}$ & $\lambda^{\delta_{5}/6}$  \\
$\rho/\tilde{\rho}$ &  $\lambda^{\delta_{5}}$ & $\lambda^{\delta_{5}}$  &$\lambda^{\delta_{5}}$ & $\lambda^{\delta_{5}}$  \\
$P/\tilde{P}$ & $\lambda^{(4/[4-\nu])\delta_{9}+(4\mu/[4-\nu])\delta_{5}}$  & $\lambda^{(4\mu/[4-\nu])\delta_{5}}$ & $\lambda^{4\delta_{5}/3}$ & $\lambda^{4\delta_{5}/3}$  \\
$T/\tilde{T}$ & $\lambda^{(1/[4-\nu])\delta_{9}+(\mu/[4-\nu])\delta_{5}}$  & $\lambda^{(\mu/[4-\nu])\delta_{5}}$ & $\lambda^{\delta_{5}/3}$ & $\lambda^{\delta_{5}/3}$ \\
$E_{r}/\tilde{E}_{r}$ & $\lambda^{(4/[4-\nu])\delta_{9}+(4\mu/[4-\nu])\delta_{5}}$   & $\lambda^{(4\mu/[4-\nu])\delta_{5}}$ & $\lambda^{4\delta_{5}/3}$ & $\lambda^{4\delta_{5}/3}$  \\
${F}_{r}/\tilde{F}_{r}$ & $\lambda^{(6/[4-\nu])\delta_{9}+([12\mu-4+\nu]/[8-2\nu])\delta_{5}}$   & $\lambda^{([12\mu-4+\nu]/[8-2\nu])\delta_{5}}$  & $\lambda^{3\delta_{5}/2}$ & $\lambda^{3\delta_{5}/2}$  \\
$P_{r}/\tilde{P}_{r}$ & $\lambda^{(4/[4-\nu])\delta_{9}+(4\mu/[4-\nu])\delta_{5}}$  & $\lambda^{(4\mu/[4-\nu])\delta_{5}}$ & $\lambda^{4\delta_{5}/3}$ & $\lambda^{4\delta_{5}/3}$\\
$\mathcal{Q}/\tilde{\mathcal{Q}}$ & $\lambda^{-\delta_{12}+([11-n]/[4-\nu])\delta_{9}+(\mu[11-n]/[4-\nu]-[m+1])\delta_{5}}$  & --- & --- & --- \\
$\kappa_{0}/\tilde{\kappa}_{0}$ & $\lambda^{\delta_{12}}$   & 1 & 1 & 1 \\
$\varepsilon_{0}/\tilde{\varepsilon}_{0}$ & $\lambda^{\delta_{9}}$   & 1 & 1 & 1  \\
\enddata
\label{Table5:tab}
\end{deluxetable}

\end{document}